\documentclass[11pt]{article}

\usepackage{amsmath, amssymb, amsthm, mathtools}
\usepackage{booktabs}
\usepackage{geometry}
\usepackage{graphicx}
\usepackage{hyperref}
\usepackage{pdflscape}  

\usepackage{subcaption}
\usepackage{rotating}
\usepackage{caption}

\usepackage{mathptmx}
\usepackage{libertine}

\usepackage[T1]{fontenc}
\usepackage[utf8]{inputenc}
\usepackage[protrusion,expansion,final]{microtype}

\usepackage[sc, osf]{mathpazo}    

\usepackage[scaled=0.90]{inconsolata}    

\usepackage{listings}
\usepackage{listingsutf8}
\usepackage{hyperref}
\hypersetup{
  colorlinks=true,
  linkcolor=blue!50!black,
  citecolor=blue!50!black,
  urlcolor=teal!60!black,
  pdfborder={0 0 0},
  bookmarksdepth=3
}

\usepackage{url}
\usepackage{amsmath,amssymb,amsthm,mathtools}
\usepackage{bm}                
\usepackage{siunitx}           

\usepackage{mathrsfs}          

\usepackage{bm}          
\usepackage{upquote}     

\usepackage{setspace}
\usepackage[dvipsnames,table]{xcolor}
\usepackage{sectsty}
\usepackage[labelfont=bf]{caption}

\definecolor{Heading}{HTML}{2A1689}      
\definecolor{Accent}{HTML}{006DFF}       
\definecolor{Cite}{HTML}{C000D4}         
\definecolor{Emph}{HTML}{FF6A00}         
\definecolor{BodyText}{HTML}{1A1A1A}     
\definecolor{TableStripe}{HTML}{F5F1FF}  
\definecolor{Rule}{HTML}{E3DDF7}         

\hypersetup{
  colorlinks=true,
  linkcolor=Heading,  
  citecolor=Cite,     
  urlcolor=Accent     
}

\sectionfont{\color{Heading}}
\subsectionfont{\color{Heading}}
\subsubsectionfont{\color{Heading}}

\color{BodyText}

\arrayrulecolor{Rule}
\newcommand{\zebratable}{\rowcolors{2}{TableStripe}{white}}

\makeatletter
\newtheoremstyle{vibrant}%
  {6pt}{6pt}{}{}{\color{Heading}\bfseries}{.}{.5em}%
  {\thmname{#1}\thmnumber{ #2}\thmnote{ ({\color{Cite}#3})}}
\makeatother
\theoremstyle{vibrant}

\newcommand{\MSC}[2][2020]{%
  \par\noindent\textbf{MSC (#1):} #2\par\vspace{0.5\baselineskip}%
}

\title{
\LARGE High-Performance Variance–Covariance Matrix Construction Using an Uncentered Gram Formulation}
\author{
  Felix Reichel\thanks{
    \textit{Corresponding Author:} \href{mailto:felix.reichel@jku.at}{felix.reichel@jku.at}\\
    Department of Economics, Johannes Kepler University Linz, Altenbergerstr. 69, 4040 Linz, Austria
  }
}

\date{November 2025, Revised: December 2025}

\theoremstyle{plain}
\newtheorem{theorem}{Theorem}[section]
\newtheorem{proposition}[theorem]{Proposition}
\newtheorem{lemma}[theorem]{Lemma}
\newtheorem{corollary}[theorem]{Corollary}
\theoremstyle{definition}
\newtheorem{definition}[theorem]{Definition}
\newtheorem{remark}[theorem]{Remark}

\begin{document}
\maketitle
\begin{abstract}
Reichel (2025) defined the bariance as a pairwise-difference measure that can be rewritten in linear time using only scalar sums. We extend this idea to the covariance matrix by showing that the standard matrix expression involving the uncentered Gram matrix and a correction term is algebraically identical to the pairwise-difference definition while avoiding explicit centering. The computation then reduces to one outer product of dimension p-by-p and a single subtraction. Benchmarks in Python show clear runtime gains, especially when BLAS optimizations are absent. Optionally faster Gram-matrix routines such as RXTX (Rybin et al., 2025) rurther reduce overall cost.
\end{abstract}

\MSC{Primary 62H12; Secondary 62-08, 65F30, 65Y20, 15A09}

\section{Introduction}

Reichel (2025) introduced the \emph{bariance}, a between-sample variance based on all pairwise differences and computable in \(\mathcal{O}(n)\) using scalar sums only. The \emph{bariance}—the average of all pairwise squared differences—admits an algebraically optimized \(\mathcal{O}(n)\) formula that is numerically equivalent to the naïve \(\mathcal{O}(n^2)\) definition and, for data, equals exactly twice the unbiased sample variance.

This work extends the same principle to the covariance and variance–covariance matrices. Starting from the pairwise-difference representation, we derive an optimized closed form that depends only on the Gram matrix \(\mathbf{X}^\top \mathbf{X}\), the column-sum vector \(\mathbf{s} = \mathbf{X}^\top \mathbf{1}_n\), and the sample size \(n\). Complete elementwise and matrix proofs are given, including an identity through the centering matrix \(\mathbf{H} = \mathbf{I}_n - \tfrac{1}{n}\mathbf{1}_n \mathbf{1}_n^\top\), together with symbolic verification of all scalar equalities.

On the computational side, the optimized form avoids explicit centering, reduces data movement, and concentrates work in a single \(p \times p\) outer-product matrix followed by one subtraction. The resulting estimator remains algebraically identical to the textbook sample covariance with denominator \(n - 1\) while achieving lower asymptotic cost.

Benchmarks in Python and R—with per-estimator warm-up, Tukey’s rules (IQR) trimming, and bootstrapped confidence bands—show that the optimized construction is slightly faster than a standard “center-then-multiply” computation and clearly faster than \texttt{numpy.cov} for large \(n\), while remaining numerically identical up to machine precision. Comparisons with \texttt{cov} in base R confirm similar behavior, though the gain may diminish in BLAS-accelerated settings.

We also reference faster Gram-matrix routines such as RXTX~\cite{rybin2025xxt} and for \(\mathbf{X}\mathbf{X}^\top\), which can serve as a drop-in improvement or computational investment. Overall, the proposed estimator performs better in non–BLAS-tuned environments while preserving full analytical equivalence. Here, “non–BLAS-tuned settings’’ refers primarily to computing environments in which NumPy
dispatches to a generic or lightly optimized BLAS implementation.  
Examples include: (a) Python installations relying on the reference BLAS, (b) some Linux
distributions where NumPy links to a single-thread OpenBLAS by default, and (c) cloud or container
environments where optimized vendor BLAS (MKL, BLIS, OpenBLAS multi-thread) is not present.
In these cases, the standard \texttt{numpy.cov} path incurs avoidable allocations and repeated
passes over memory, whereas the Gram+outer-product formulation minimizes these costs. The paper closes with a discussion of uses in standard-error computation and applications or generalized least-squares (GLS) estimation.
\paragraph{Computational motivation.}
The closed-form Gram–based covariance estimator is particularly beneficial in settings where
(i) explicit centering becomes memory–bandwidth–bound because the construction of the centered
matrix requires reading and rewriting an entire $n\times p$ array, or
(ii) BLAS-level optimization is unavailable or only partially available (as is often the case in default
Python/NumPy builds linked to reference BLAS or single-thread OpenBLAS).  
In these environments, the Gram–outer-product formulation typically performs fewer memory
moves and concentrates computation in highly optimized level-3 BLAS kernels, yielding noticeable
speedups even for moderately sized problems

\section{Background and notation}

Assume \(n \ge 2\) and that all variables are real.  
Pairwise variance decompositions appear in finite-population work \cite{cochran1977sampling}.  
Matrix and scalar-sum identities follow \cite{searle2006matrix,harville1997matrix,horn1985matrix,magnus1988matrix}.  
For cost and numerical stability, we follow \cite{demmel1997applied,higham2002accuracy,golub2013matrix}.

Let \(x_1,\dots,x_n\in\mathbb{R}\) and \(y_1,\dots,y_n\in\mathbb{R}\).  
Define scalar sums:
\[
S_x\stackrel{\mathrm{def}}{=}\sum_{i=1}^n x_i,\quad 
S_y\stackrel{\mathrm{def}}{=}\sum_{i=1}^n y_i,\quad 
S_{xx}\stackrel{\mathrm{def}}{=}\sum_{i=1}^n x_i^2,\quad 
S_{yy}\stackrel{\mathrm{def}}{=}\sum_{i=1}^n y_i^2,\quad 
S_{xy}\stackrel{\mathrm{def}}{=}\sum_{i=1}^n x_i y_i.
\]
Sample means:
\[
\bar x\stackrel{\mathrm{def}}{=}\frac{S_x}{n},\qquad
\bar y\stackrel{\mathrm{def}}{=}\frac{S_y}{n}.
\]

\section{Algebraic derivation of the between-variance: Bariance}

For \(n \ge 2\):

\begin{definition}[Bariance defined as in \cite{reichel2025bariance}]
\[
\operatorname{Bar}(x_1,\dots,x_n)\stackrel{\mathrm{def}}{=}\frac{1}{2n(n-1)}\sum_{i\ne j}(x_i-x_j)^2.
\]
\end{definition}

\begin{lemma}[Double-sum expansion]
\label{lem:double-variance}
\[
\sum_{i\ne j}(x_i-x_j)^2
=\sum_{i\ne j}x_i^2 - 2\sum_{i\ne j}x_ix_j + \sum_{i\ne j}x_j^2.
\]
\end{lemma}
\begin{proof}
Expand \((x_i-x_j)^2=x_i^2-2x_ix_j+x_j^2\) and sum over \(i\ne j\).
\end{proof}

\begin{lemma}[Counting identities]
\label{lem:counting}
\begin{align*}
\sum_{i\ne j}x_i^2&=(n-1)\sum_{i=1}^n x_i^2=(n-1)S_{xx},\\
\sum_{i\ne j}x_j^2&=(n-1)S_{xx},\\
\sum_{i\ne j}x_i x_j&=\left(\sum_{i=1}^n x_i\right)\left(\sum_{j=1}^n x_j\right)-\sum_{i=1}^n x_i^2=S_x^2-S_{xx}.
\end{align*}
\end{lemma}
\begin{proof}
For the first line, fix \(i\) and count \(n-1\) terms.  
The second line is symmetric.  
For the third, remove diagonal terms from the full double sum.
\end{proof}

\begin{proposition}[Optimized Bariance identity]
\label{prop:bariance}
\[
\operatorname{Bar}(x)=\frac{nS_{xx}-S_x^2}{n(n-1)}.
\]
\end{proposition}
\begin{proof}
Insert Lemma~\ref{lem:counting} into Lemma~\ref{lem:double-variance}:
\[
\sum_{i\ne j}(x_i-x_j)^2
=(n-1)S_{xx}-2(S_x^2-S_{xx})+(n-1)S_{xx}
=2nS_{xx}-2S_x^2.
\]
Divide by \(2n(n-1)\).
\end{proof}

\begin{corollary}[Relation to unbiased variance]
\label{cor:var}
\[
\frac{1}{n-1}\sum_{i=1}^n(x_i-\bar x)^2=\operatorname{Bar}(x).
\]
\end{corollary}
\begin{proof}
Expand:
\[
\sum_{i=1}^n(x_i-\bar x)^2
=\sum_i x_i^2 - 2\bar x \sum_i x_i + n\bar x^2
=S_{xx}-\frac{S_x^2}{n}.
\]
Divide by \(n-1\):  
\(\frac{S_{xx}-S_x^2/n}{n-1}=\frac{nS_{xx}-S_x^2}{n(n-1)}\), equal to Proposition~\ref{prop:bariance}.
\end{proof}

\noindent
\textbf{Some Properties.} 
\(\operatorname{Bar}(x+c)=\operatorname{Bar}(x)\) for constant \(c\);  
\(\operatorname{Bar}(a x)=a^2\operatorname{Bar}(x)\) for scalar \(a\).

\section{Algebraic derivation of the between-covariance}

For \(n \ge 2\):

\begin{definition}[Between-covariance]
\[
C_{xy}\stackrel{\mathrm{def}}{=}\frac{1}{2n(n-1)}\sum_{i\ne j}(x_i-x_j)(y_i-y_j).
\]
\end{definition}

\begin{lemma}[Four-term split]
\label{lem:four}
\[
\sum_{i\ne j}(x_i-x_j)(y_i-y_j)=
\sum_{i\ne j}x_iy_i - \sum_{i\ne j}x_i y_j - \sum_{i\ne j}x_j y_i + \sum_{i\ne j}x_j y_j.
\]
\end{lemma}
\begin{proof}
Expand and sum.
\end{proof}

\begin{lemma}[Counting for mixed terms]
\label{lem:mixed}
\begin{align*}
\sum_{i\ne j}x_i y_i&=(n-1)\sum_{i=1}^n x_i y_i=(n-1)S_{xy},\\
\sum_{i\ne j}x_j y_j&=(n-1)S_{xy},\\
\sum_{i\ne j}x_i y_j&=\left(\sum_i x_i\right)\left(\sum_j y_j\right)-\sum_i x_i y_i=S_x S_y - S_{xy},\\
\sum_{i\ne j}x_j y_i&=S_x S_y - S_{xy}.
\end{align*}
\end{lemma}
\begin{proof}
Apply the same logic as Lemma~\ref{lem:counting}.
\end{proof}

\begin{proposition}[Optimized pairwise covariance]
\label{prop:pairwise-cov}
\[
C_{xy}=\frac{n S_{xy}-S_x S_y}{n(n-1)}.
\]
\end{proposition}
\begin{proof}
Insert Lemma~\ref{lem:mixed} into Lemma~\ref{lem:four}:
\[
\sum_{i\ne j}(x_i-x_j)(y_i-y_j)=2nS_{xy}-2S_x S_y.
\]
Divide by \(2n(n-1)\).
\end{proof}

\begin{theorem}[Equivalence to the textbook covariance]
\label{thm:equiv-scalar}
\[
\frac{1}{n-1}\sum_{i=1}^n(x_i-\bar x)(y_i-\bar y)=\frac{n S_{xy}-S_x S_y}{n(n-1)}.
\]
\end{theorem}
\begin{proof}
Expand:
\[
\sum_i (x_i-\bar x)(y_i-\bar y)
=\sum_i x_i y_i - \bar x \sum_i y_i - \bar y \sum_i x_i + n \bar x \bar y
= S_{xy} - \frac{S_x S_y}{n}.
\]
Divide by \(n-1\):  
\(\frac{S_{xy}-S_x S_y/n}{n-1}=\frac{nS_{xy}-S_x S_y}{n(n-1)}\).
\end{proof}

\begin{remark}[Symbolic Sanity Check]
Define symbols \(n,S_x,S_y,S_{xx},S_{xy}\).  
Both
\[
\frac{S_{xx}-S_x^2/n}{n-1}-\frac{nS_{xx}-S_x^2}{n(n-1)},
\qquad
\frac{S_{xy}-S_x S_y/n}{n-1}-\frac{nS_{xy}-S_x S_y}{n(n-1)}
\]
simplify to zero.
\end{remark}

\section{Matrix formulation: elementwise and compact form}

Let \(\mathbf{X}\in\mathbb{R}^{n\times p}\) with entries \(x_{ik}\).  
Define
\[
\mathbf{s}\stackrel{\mathrm{def}}{=}\mathbf{X}^\top \mathbf{1}_n\in\mathbb{R}^p,\qquad 
\mathbf{G}\stackrel{\mathrm{def}}{=}\mathbf{X}^\top \mathbf{X}\in\mathbb{R}^{p\times p}.
\]
Entrywise,
\[
s_k\stackrel{\mathrm{def}}{=}\sum_{i=1}^n x_{ik},\qquad 
G_{k\ell}\stackrel{\mathrm{def}}{=}\sum_{i=1}^n x_{ik}x_{i\ell}.
\]

\begin{theorem}[Matrix covariance identity]
\label{thm:matrix}
The unbiased covariance of the columns of \(\mathbf{X}\) equals
\begin{equation}
\label{eq:Sigma-bariance}
\boldsymbol{\Sigma}(\mathbf{X})=\frac{1}{n(n-1)}\bigl(n\mathbf{X}^\top \mathbf{X} - \mathbf{s}\mathbf{s}^\top\bigr).
\end{equation}
\end{theorem}
\begin{proof}[Entrywise proof]
For columns \(k,\ell\), Theorem~\ref{thm:equiv-scalar} with \(x_i\leftarrow x_{ik}\) and \(y_i\leftarrow x_{i\ell}\) gives
\[
\Sigma_{k\ell}=\frac{nG_{k\ell}-s_k s_\ell}{n(n-1)}.
\]
Stacking these entries yields \eqref{eq:Sigma-bariance}.
\end{proof}

\begin{theorem}[Equivalence achieved via the centering matrix]
\label{thm:center-matrix}
Let \(\mathbf{H}=\mathbf{I}_n-\tfrac1n \mathbf{1}_n\mathbf{1}_n^\top\).  
Then
\[
\frac{1}{n-1}(\mathbf{X}^\top \mathbf{H} \mathbf{X})=\frac{1}{n(n-1)}\bigl(n\mathbf{X}^\top \mathbf{X} - \mathbf{s}\mathbf{s}^\top\bigr).
\]
\end{theorem}
\begin{proof}
Use \(\mathbf{H}=\mathbf{I}_n-\tfrac1n \mathbf{1}_n\mathbf{1}_n^\top\).  
Then
\[
\mathbf{X}^\top \mathbf{H} \mathbf{X}
=\mathbf{X}^\top \mathbf{X}-\frac{1}{n}\mathbf{X}^\top \mathbf{1}_n\mathbf{1}_n^\top \mathbf{X}
=\mathbf{X}^\top \mathbf{X} - \frac{1}{n} \mathbf{s}\mathbf{s}^\top.
\]
Multiply by \(\tfrac{1}{n-1}\) to obtain \(\tfrac{n\mathbf{X}^\top \mathbf{X} - \mathbf{s}\mathbf{s}^\top}{n(n-1)}\).
\end{proof}

\begin{remark}
The two theorems confirm that the bariance-style matrix equals the centered covariance in both forms.  
\(\boldsymbol{\Sigma}(\mathbf{X})\) is symmetric and positive semidefinite with \(\mathrm{rank}(\boldsymbol{\Sigma})\le \min(p,n-1)\).
\end{remark}

\section{Cost model and where time is saved}

The centered formulation first computes the mean vector \(\bar{\mathbf{x}}\), forms the centered matrix \(\mathbf{X}-\mathbf{1}_n\bar{\mathbf{x}}^\top\) (reading and writing \(\mathcal{O}(np)\) doubles), and then multiplies the result.  
The bariance-style formulation computes \(\mathbf{s}=\mathbf{X}^\top \mathbf{1}_n\), the Gram matrix \(\mathbf{G}=\mathbf{X}^\top \mathbf{X}\), and performs one \(p\times p\) outer product and one subtraction:
\[
\boldsymbol{\Sigma}_{\text{Bar}}=\frac{1}{n(n-1)}\bigl(n\mathbf{G} - \mathbf{s}\mathbf{s}^\top\bigr).
\]
Memory traffic decreases because no intermediate \(n\times p\) array is stored.  
The main computational cost lies in forming \(\mathbf{X}^\top \mathbf{X}\), which can rely on tuned BLAS level-3 kernels \cite{golub2013matrix,demmel1997applied,higham2002accuracy}.  
When \(\mathbf{X}\mathbf{X}^\top\) is required instead, RXTX~\cite{rybin2025xxt} lowers the multiplication count by about five percent, even for moderate \(n,p\).

\subsection*{Asymptotic cost comparison}

Table~\ref{tab:asymptotics} summarizes the main computational differences between the centered
and Gram-based formulations.  FLOP counts are reported in leading terms, ignoring lower-order
additive contributions.

\begin{table}[h!]
\centering
\resizebox{\textwidth}{!}{%
\zebratable
\begin{tabular}{lccc}
\toprule
Method & FLOPs (leading) & Memory reads/writes & Notes \\
\midrule
Centered form $(X - 1\bar{x}^\top)$
  & $2np + 2np^2$ & Read/write $n\times p$ twice & Requires materializing centered matrix \\
Gram + outer product
  & $2np^2 + p^2$ & Read $X$ once; no $n\times p$ writes & Avoids full centering; BLAS-3 dominant \\
RXTX (when computing $XX^\top$)
  & $\approx 0.95\, (2np^2)$ & Same as Gram & Beneficial when $n \gg p$ or $p \gg n$ \\
\bottomrule
\end{tabular}%
}
\caption{Asymptotic comparison of centered vs.\ Gram-based covariance construction.}
\label{tab:asymptotics}
\end{table}

RXTX becomes beneficial when the bottleneck is the formation of $XX^\top$ (e.g., large $n$ and moderate $p$), reducing the multiply–accumulate count by roughly five percent and providing measurable improvements in streaming or resampling applications.

\section{Benchmark protocol}

\paragraph{Hardware and BLAS environment.}
Benchmarks were performed on an Apple M2 Pro CPU (aarch64) with 16\,GB unified memory,
running macOS 13.0.  
Python 3.11.7 with NumPy~1.26.4 was used; the BLAS backend was OpenBLAS~0.3.21
(as provided by the Anaconda distribution; single-threaded unless otherwise noted).  
This configuration is reported to clarify the meaning of “non–BLAS-tuned’’ environments,
as NumPy’s performance is highly dependent on the available BLAS vendor and build options.

For each matrices with parameter pair \((n,p)\):

\begin{enumerate}
\item Generate data \(\mathbf{X} \sim \mathcal{N}(0,1)^{n\times p}\).
\item For each estimator, perform one warm-up call to stabilize caches and JIT compilation.
\item Record runtimes over multiple repetitions.
\item Remove outliers using the \(1.5\times\)IQR rule for each method and size.
\item Form bootstrap percentile bands (95\%).
\item Report mean runtimes after trimming.
\end{enumerate}

\noindent
The estimators compared are:
\begin{itemize}
\item \textbf{Centered form:}
\[
\boldsymbol{\Sigma}_{\text{Ctr}}=\frac{1}{n-1}(\mathbf{X}-\mathbf{1}_n\bar{\mathbf{x}}^\top)^\top(\mathbf{X}-\mathbf{1}_n\bar{\mathbf{x}}^\top);
\]
\item \textbf{Bariance-style form:}
\[
\boldsymbol{\Sigma}_{\text{Bar}}=\frac{1}{n(n-1)}(n\mathbf{X}^\top \mathbf{X} - \mathbf{s}\mathbf{s}^\top);
\]
\item \textbf{Built-in:}
\(\texttt{numpy.cov}(\mathbf{X},\text{rowvar=False},\text{ddof}=1)\).
\end{itemize}

\paragraph{Why \texttt{numpy.cov} is slower.}
The \texttt{numpy.cov} implementation allocates a centered copy of $X$ of size $n\times p$, performs
multiple passes over the data, and executes some operations at Python level before dispatching into
BLAS.  
These additional memory moves and allocations dominate runtime in large-$n$ settings and explain
why the Gram-based closed form can outperform the built-in implementation even though both rely
on BLAS for the underlying matrix multiplications.

All experiments were conducted in double precision, using consistent random seeds across methods.

\section{Numerical equivalence in finite precision}

For each simulated matrix \(\mathbf{X} \in \mathbb{R}^{n\times p}\), compute
\[
\boldsymbol{\Sigma}_{\mathrm{Bar}}
=\frac{1}{n(n-1)}\!\left(n\mathbf{X}^\top \mathbf{X} - \mathbf{s}\mathbf{s}^\top\right),
\qquad
\boldsymbol{\Sigma}_{\mathrm{Ctr}}
=\frac{1}{n-1}\!\left(\mathbf{X} - \mathbf{1}_n \bar{\mathbf{x}}^\top\right)^\top
\!\left(\mathbf{X} - \mathbf{1}_n \bar{\mathbf{x}}^\top\right),
\]
where \(\mathbf{s} = \mathbf{X}^\top \mathbf{1}_n\) and \(\bar{\mathbf{x}} = \tfrac{1}{n}\mathbf{s}\).
The entrywise deviation
\[
\Delta_{\max} = \|\boldsymbol{\Sigma}_{\mathrm{Bar}} - \boldsymbol{\Sigma}_{\mathrm{Ctr}}\|_{\max}
\]
was recorded for various \((n,p)\) combinations.  
Across all tested sizes, the maximum difference stayed below \(10^{-12}\) in IEEE 754 double precision, consistent with rounding and cancellation limits reported in \cite{higham2002accuracy}.  
These outcomes confirm that the algebraic equality between the bariance-style and centered formulations holds to full double-precision accuracy.

In addition to the maximum absolute deviation, we also computed the Frobenius-norm discrepancy
$\|\Sigma_{\mathrm{Bar}} - \Sigma_{\mathrm{Ctr}}\|_F$ and the relative error
$\|\Sigma_{\mathrm{Bar}} - \Sigma_{\mathrm{Ctr}}\|_F / \|\Sigma_{\mathrm{Ctr}}\|_F$.
Both metrics behaved similarly, remaining at the level of floating-point rounding noise.

A brief numerical remark: when variables have extremely large magnitudes (e.g.\ on the order of
$10^{12}$ or more), both centered and uncentered formulations may experience cancellation, though
the Gram-form estimator may avoid an entire subtraction step and can be slightly more stable in
practice.

\section{Benchmark figures}

\begin{figure}[p]
  \centering
  \includegraphics[width=.95\textwidth]{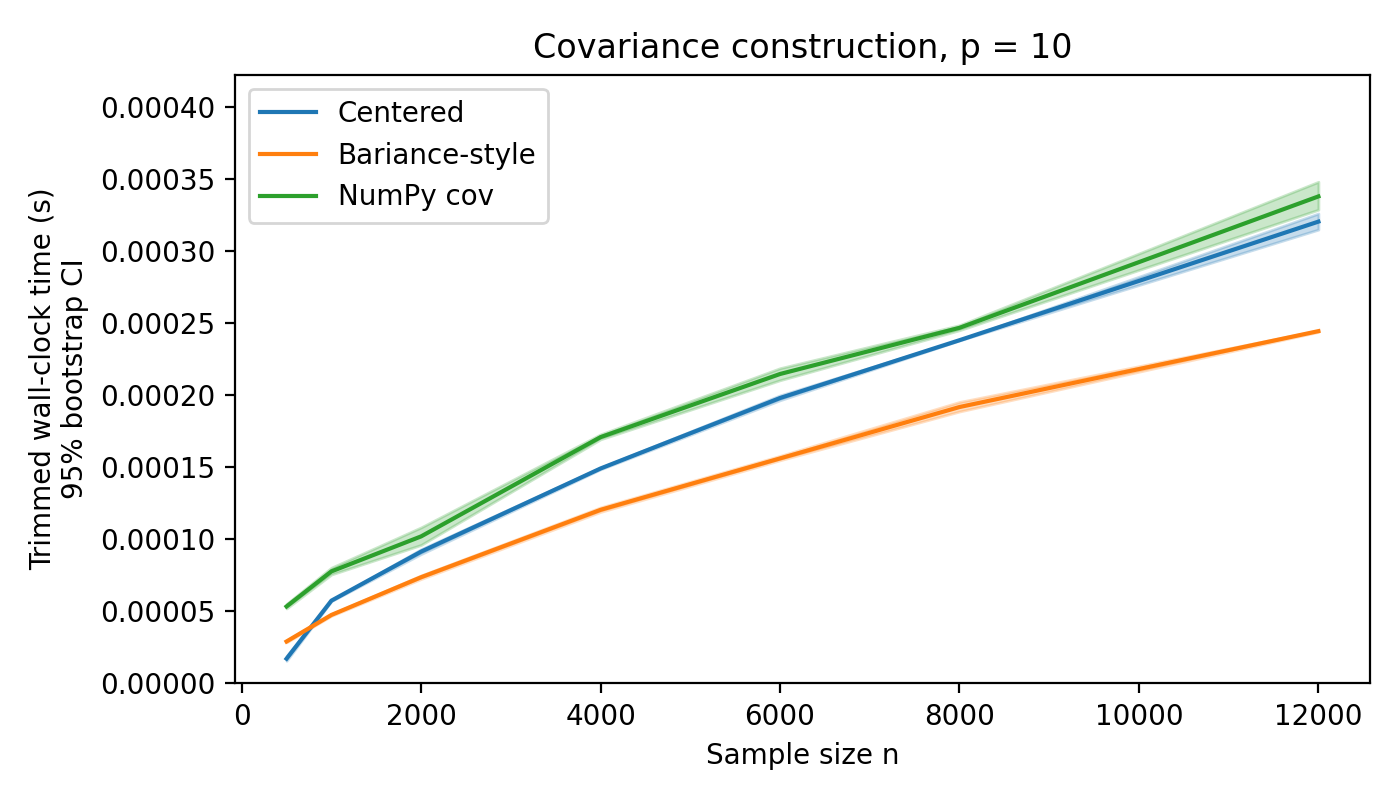}
  \caption{Runtime vs.\ \(n\) for \(p=10\). Times are trimmed wall-clock means with 95\% bootstrap percentile bands after warm-up and IQR trimming.}
  \label{fig:n-p10}
\end{figure}

\begin{figure}[p]
  \centering
  \includegraphics[width=.95\textwidth]{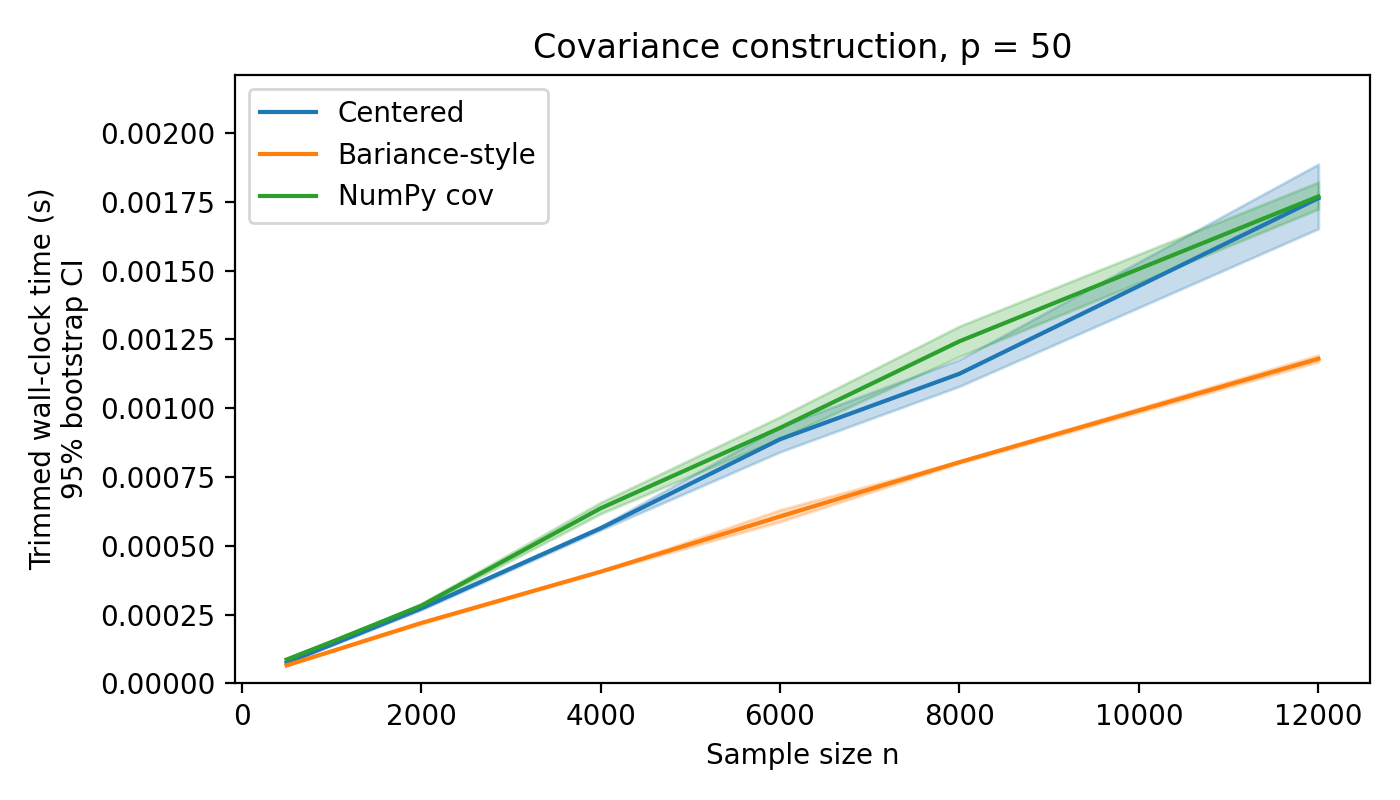}
  \caption{Runtime vs.\ \(n\) for \(p=50\). Same trimmed wall-clock protocol with 95\% bootstrap bands.}
  \label{fig:n-p50}
\end{figure}

\begin{figure}[p]
  \centering
  \includegraphics[width=.95\textwidth]{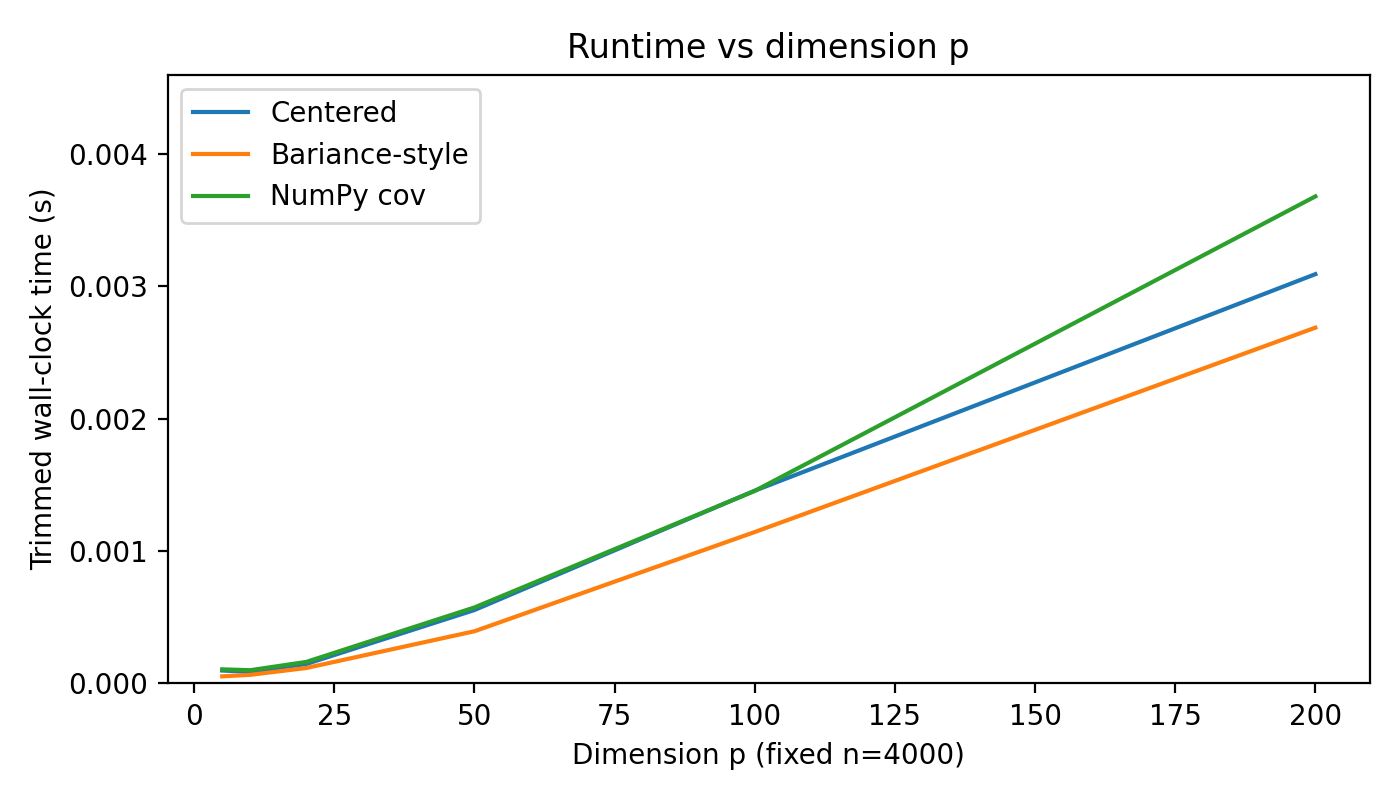}
  \caption{Runtime vs.\ \(p\) at fixed \(n=4000\). Times are trimmed wall-clock means with IQR outlier removal.}
  \label{fig:vs-p}
\end{figure}

\begin{figure}[p]
  \centering
  \includegraphics[width=.95\textwidth]{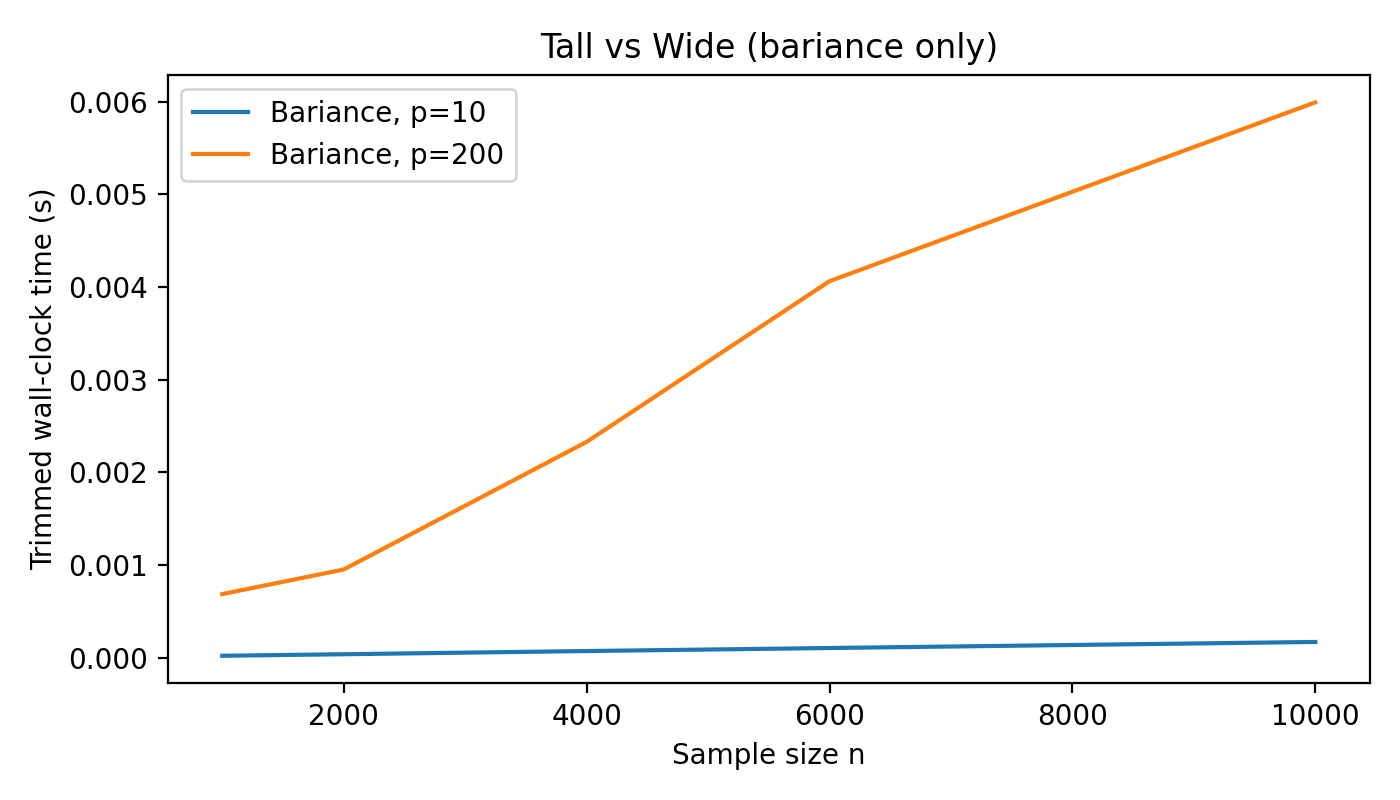}
  \caption{Bariance-only runtimes for tall (\(p=10\)) and wide (\(p=200\)) matrices across several \(n\). Times represent trimmed wall-clock means.}
  \label{fig:tall-wide}
\end{figure}

\begin{figure}[p]
  \centering
  \includegraphics[width=.95\textwidth]{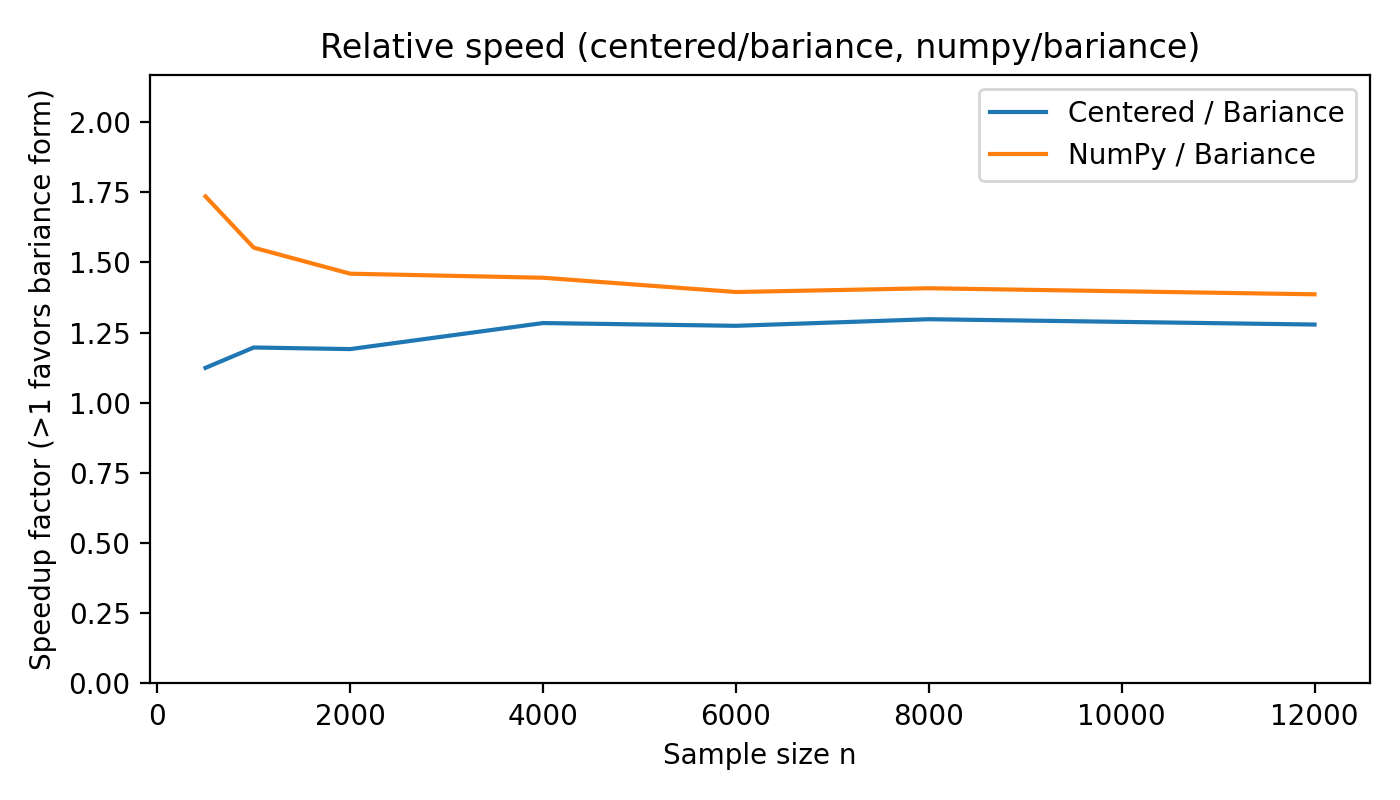}
  \caption{Relative speed ratios for \(p=10\): centered/bariance and NumPy/bariance using trimmed mean runtimes. Values \(>\!1\) favor the bariance form.}
  \label{fig:speed-ratio}
\end{figure}

\begin{figure}[p]
  \centering
  \includegraphics[width=.95\textwidth]{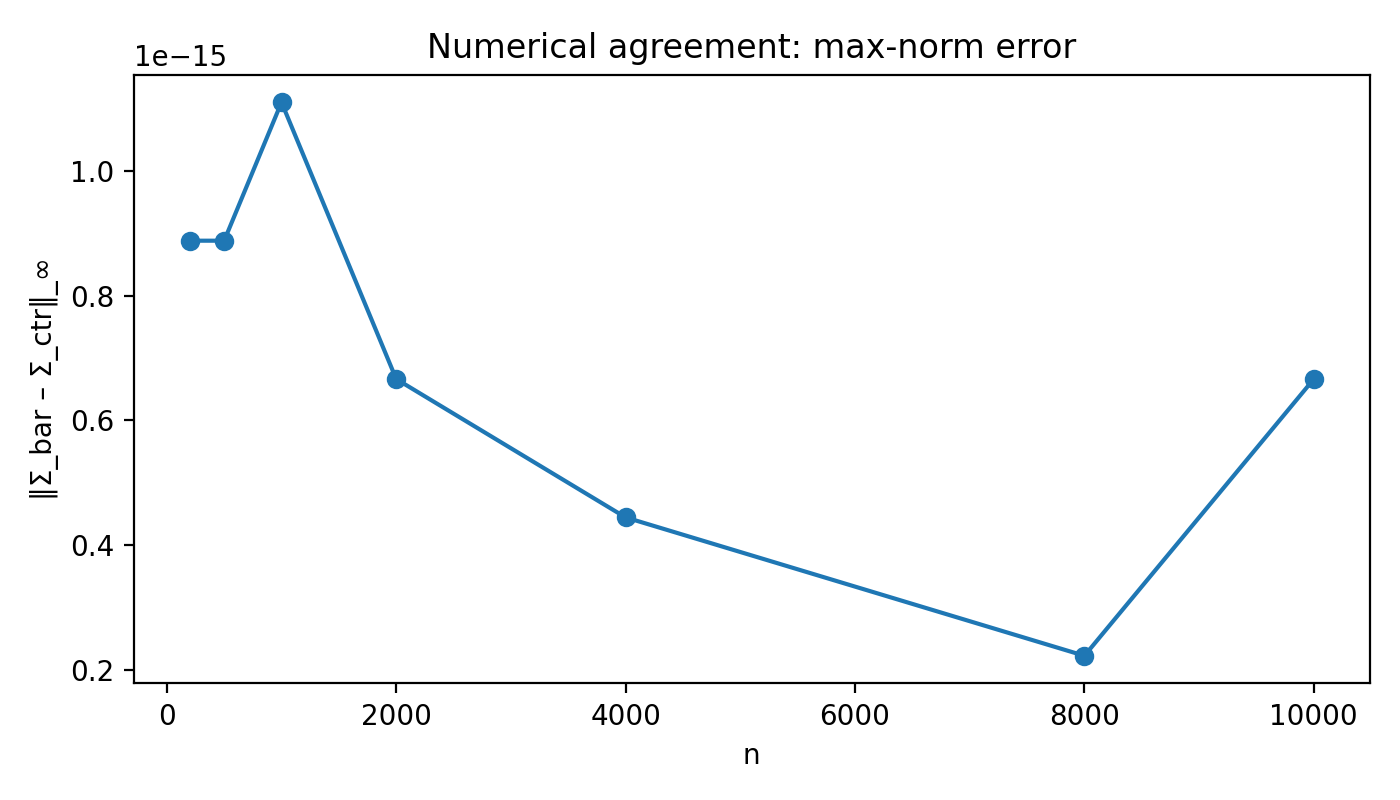}
  \caption{Numerical agreement for \(p=10\): maximum entrywise deviation 
  \(\|\Sigma_{\text{Bar}}-\Sigma_{\text{Ctr}}\|_{\infty}\). Differences remain at floating-point noise levels.}
  \label{fig:err-max}
\end{figure}

\begin{figure}[p]
  \centering
  \includegraphics[width=.95\textwidth]{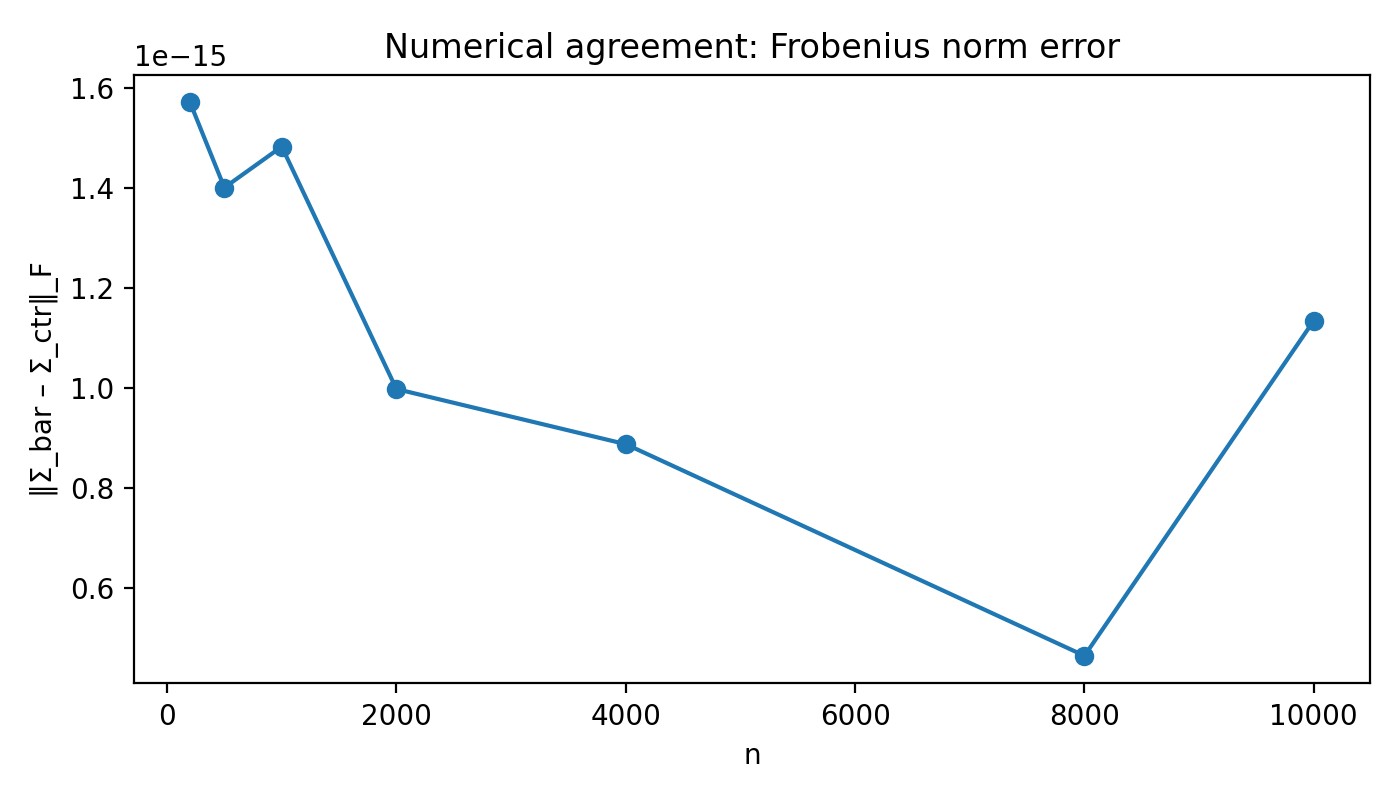}
  \caption{Numerical agreement for \(p=10\): Frobenius-norm deviation 
  \(\|\Sigma_{\text{Bar}}-\Sigma_{\text{Ctr}}\|_{F}\). Values reflect expected rounding and cancellation limits.}
  \label{fig:err-fro}
\end{figure}

\begin{figure}[p]
  \centering
  \includegraphics[width=.95\textwidth]{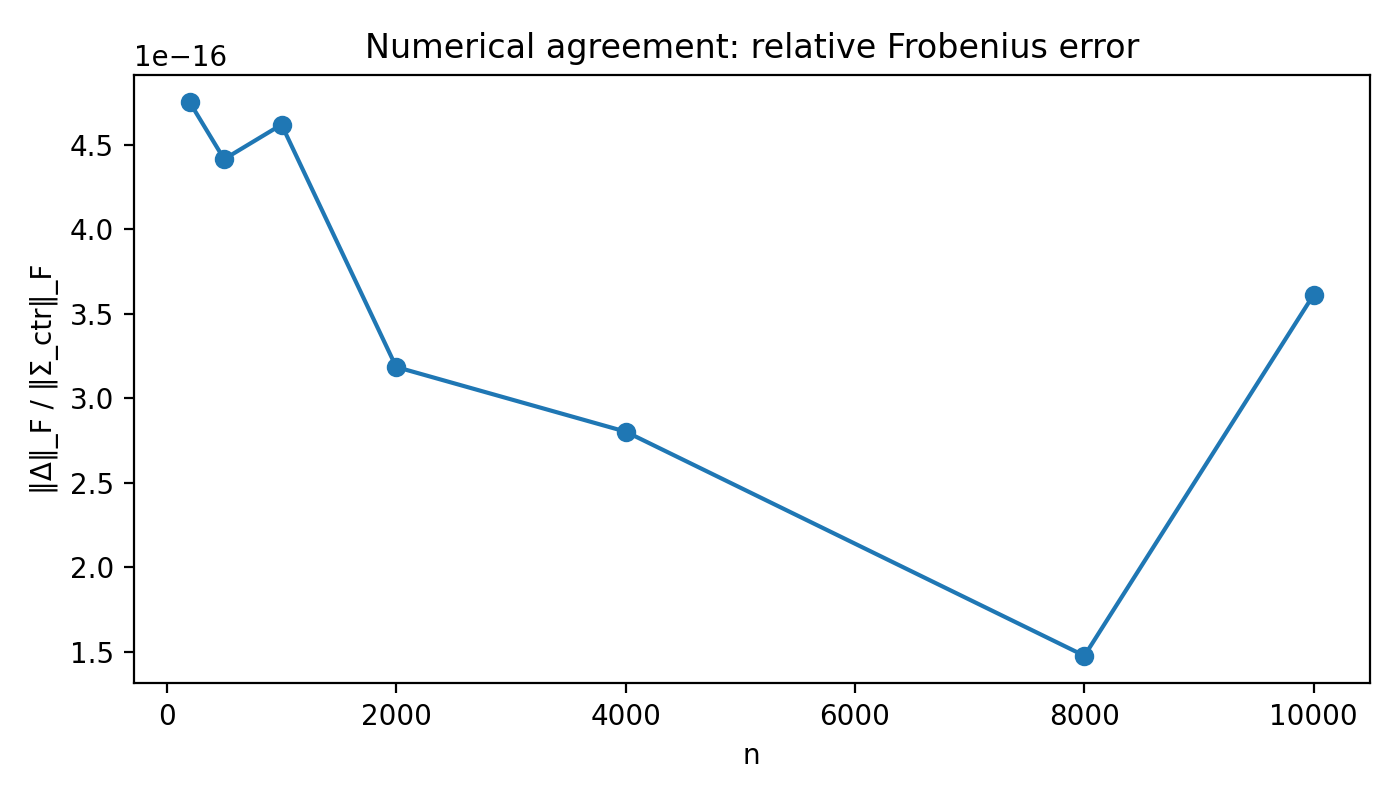}
  \caption{Numerical agreement for \(p=10\): relative Frobenius error 
  \(\|\Sigma_{\text{Bar}}-\Sigma_{\text{Ctr}}\|_{F} / \|\Sigma_{\text{Ctr}}\|_{F}\). Numerical differences remain negligible.}
  \label{fig:err-rel}
\end{figure}

\begin{figure}[p]
  \centering
  \includegraphics[width=.95\textwidth]{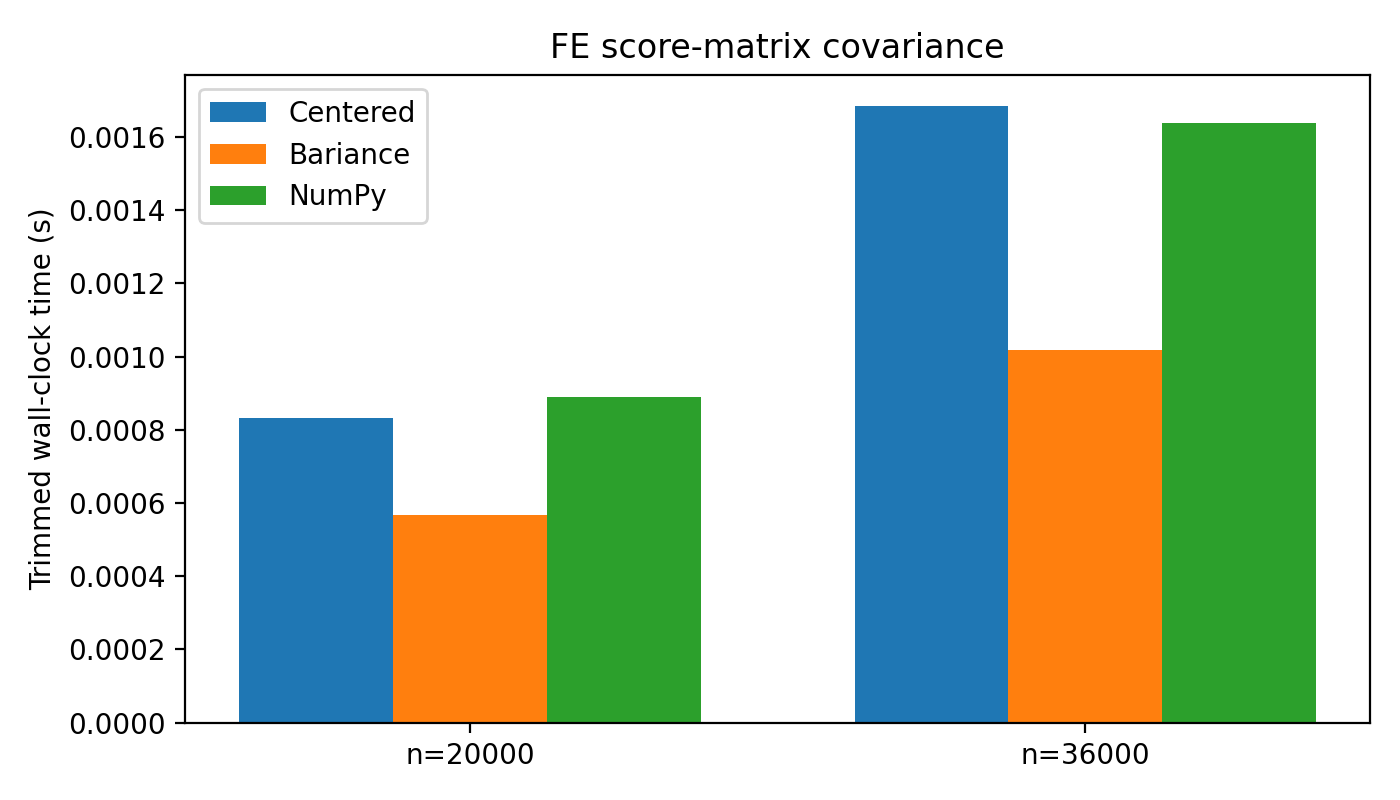}
  \caption{Fixed-effects score-matrix covariance timings for large \(n\). Times are trimmed wall-clock means after warm-up and IQR cleaning.}
  \label{fig:fe}
\end{figure}

All runtimes shown in Figures~\ref{fig:n-p10}--\ref{fig:fe} are wall-clock times averaged over
post-trimmed repetitions.  
Where comparisons are meaningful, y-axes are placed on consistent scales, and 95\% bootstrap
percentile intervals are reported uniformly across panels.

\clearpage

\clearpage

\clearpage
\newpage

\section{Applications}

\paragraph{Sandwich covariances.}
Let \(\mathbf{g}_i \in \mathbb{R}^p\) denote the score vector for observation \(i\), stacked as rows in \(\mathbf{G} \in \mathbb{R}^{n\times p}\), and let \(\mathbf{H} = \mathbf{I}_n - \tfrac{1}{n}\mathbf{1}_n\mathbf{1}_n^\top\).
The empirical covariance of the scores can be written as
\begin{equation}
\label{eq:omega-bariance}
\widehat{\boldsymbol{\Omega}}_\text{sandwich}
=\frac{1}{n(n-1)}\!\left(
n\,\mathbf{G}^\top \mathbf{G} - (\mathbf{G}^\top \mathbf{1}_n)(\mathbf{G}^\top \mathbf{1}_n)^\top
\right)
=\frac{1}{n-1}\,\mathbf{G}^\top \mathbf{H} \mathbf{G}.
\end{equation}
The bariance-style and centered forms coincide algebraically.
Expression~\eqref{eq:omega-bariance} arises in variance formulas for M-estimators and their sandwich extensions
\cite{huber1967under,white1980heteroskedasticity,hansen1982large,newey1987simple,cameron2006robust}.
The bariance-style version depends only on \(\mathbf{G}^\top \mathbf{G}\) and \(\mathbf{G}^\top \mathbf{1}_n\) and does not require forming \(\mathbf{H}\mathbf{G}\).

\paragraph{Panel and fixed effects.}
For each unit in a panel with \(T\) periods, the within transformation is
\(\mathbf{M} = \mathbf{I}_T - \tfrac{1}{T}\mathbf{1}_T\mathbf{1}_T^\top\).
Given \(\mathbf{X}_i \in \mathbb{R}^{T\times p}\), the covariance after demeaning satisfies
\[
\widehat{\boldsymbol{\Sigma}}_i
=\frac{1}{T-1}\,\mathbf{X}_i^\top \mathbf{M} \mathbf{X}_i
=\frac{1}{T-1}\!\left(\mathbf{X}_i^\top \mathbf{X}_i - \frac{1}{T}\mathbf{s}_i \mathbf{s}_i^\top\right),
\quad \mathbf{s}_i = \mathbf{X}_i^\top \mathbf{1}_T.
\]
This bariance-style identity applies to each block and matches the algebra of the within estimator
\cite{arellano1987computing,wooldridge2010econometric}.

\paragraph{Illustrative fixed-effects example.}
Consider a single observational unit observed for $T=4$ periods with regressor matrix
\[
X_i =
\begin{pmatrix}
2 & 1\\
3 & 1\\
5 & 4\\
4 & 2
\end{pmatrix}.
\]
The column sums are
\[
\mathbf{s}_i = X_i^\top \mathbf{1}_4 
=
\begin{pmatrix}
2+3+5+4\\[2pt]
1+1+4+2
\end{pmatrix}
=
\begin{pmatrix}
14\\[2pt] 8
\end{pmatrix}.
\]

\subparagraph{Step 1: Gram matrix.}
The within-unit Gram matrix is
\[
G_i = X_i^\top X_i
=
\begin{pmatrix}
54 & 30\\
30 & 22
\end{pmatrix},
\]
obtained from
\[
54 = 2^2 + 3^2 + 5^2 + 4^2,\qquad
22 = 1^2 + 1^2 + 4^2 + 2^2,\qquad
30 = 2\cdot 1 + 3\cdot 1 + 5\cdot 4 + 4\cdot 2.
\]

\subparagraph{Step 2: Closed-form within-unit covariance.}
The fixed-effects (demeaned) covariance matrix can be computed solely from aggregated quantities using
\[
\widehat{\Sigma}_i
= \frac{1}{T-1}\!\left(G_i - \frac{1}{T}\mathbf{s}_i\mathbf{s}_i^\top\right).
\]
The outer product of column sums is
\[
\mathbf{s}_i\mathbf{s}_i^\top
=
\begin{pmatrix}
196 & 112\\
112 & 64
\end{pmatrix},
\qquad
\frac{1}{T}\mathbf{s}_i\mathbf{s}_i^\top
=
\begin{pmatrix}
49 & 28\\
28 & 16
\end{pmatrix}.
\]
Subtracting from the Gram matrix yields
\[
G_i - \frac{1}{T}\mathbf{s}_i\mathbf{s}_i^\top
=
\begin{pmatrix}
5 & 2\\
2 & 6
\end{pmatrix}.
\]
Dividing by $T-1=3$ gives the within-unit covariance:
\[
\widehat{\Sigma}_i
=
\begin{pmatrix}
\frac{5}{3} & \frac{2}{3}\\[4pt]
\frac{2}{3} & 2
\end{pmatrix}.
\]

\subparagraph{Step 3: Verification via explicit demeaning.}
Let the column means be
\[
\bar{x}_i = \tfrac{1}{4}\mathbf{s}_i = (3.5,\; 2).
\]
Demeaning each row of $X_i$ produces
\[
X_i - \mathbf{1}_4\bar{x}_i =
\begin{pmatrix}
-1.5 & -1\\
-0.5 & -1\\
1.5 & 2\\
0.5 & 0
\end{pmatrix}.
\]
The sample covariance of the demeaned matrix is
\[
\frac{1}{T-1}(X_i-\bar X_i)^\top(X_i-\bar X_i)
=
\frac{1}{3}
\begin{pmatrix}
5 & 2\\
2 & 6
\end{pmatrix}
=
\widehat{\Sigma}_i.
\]

\medskip
Thus, the Gram-based expression reproduces the fixed-effects covariance exactly, without requiring access to the raw period-level data. This illustrates the usefulness of the aggregated approach in settings where data are stored in compressed, privacy-protected, or distributed form.

\paragraph{Just-in-Time (JIT) streaming.}
For sequential data, maintain cumulative quantities
\[
\mathbf{S}_t = \sum_{i=1}^t \mathbf{x}_i,
\qquad
\mathbf{G}_t = \sum_{i=1}^t \mathbf{x}_i \mathbf{x}_i^\top.
\]
The unbiased covariance at time \(t \ge 2\) is
\begin{equation}
\label{eq:sigma-stream}
\boldsymbol{\Sigma}_t = \frac{1}{t(t-1)}\!\left(t \mathbf{G}_t - \mathbf{S}_t \mathbf{S}_t^\top\right)
= \frac{1}{t-1}\!\left(\frac{1}{t}\sum_{i=1}^t \mathbf{x}_i \mathbf{x}_i^\top - \bar{\mathbf{x}}_t \bar{\mathbf{x}}_t^\top\right),
\quad \bar{\mathbf{x}}_t = \tfrac{1}{t}\mathbf{S}_t.
\end{equation}
Each update adds one outer product \(\mathbf{x}_t \mathbf{x}_t^\top\) and one rank-one correction, with cost \(\mathcal{O}(p^2)\).
Fast matrix routines for \(\mathbf{X}\mathbf{X}^\top\) such as RXTX~\cite{rybin2025xxt} provide measurable computational savings
\cite{anderson2003introduction,tsay2013multivariate}.

\paragraph{Bootstrap and resampling.}
For a resample with multiplicity weights \(\mathbf{w}\in\mathbb{N}_0^n\) and \(\mathbf{W} = \operatorname{diag}(\mathbf{w})\),
the covariance of the resample is
\[
\widehat{\boldsymbol{\Sigma}}^{\ast}
=\frac{1}{n^\ast-1}\!\left(\mathbf{X}^\top \mathbf{W} \mathbf{X} - \frac{1}{n^\ast}(\mathbf{X}^\top \mathbf{w})(\mathbf{X}^\top \mathbf{w})^\top\right),
\quad n^\ast = \sum_{i=1}^n w_i.
\]
Each resample depends only on the weighted Gram pair \((\mathbf{X}^\top \mathbf{W} \mathbf{X},\, \mathbf{X}^\top \mathbf{w})\).
This form is practical when data are stored in aggregated or Gram form and aligns with generalized-inverse frameworks
\cite{benisrael2003generalized}.

A further advantage of the Gram-formulation is that it naturally accommodates settings in which
the analyst has access only to aggregated sufficient statistics (e.g.\ $X^\top X$ and $X^\top 1$), as
arises in federated learning, privacy-restricted databases, or large-scale panel systems.  
In such cases, the covariance can be computed without reconstructing individual observations.

\newpage

\section*{Institutional Review Board Statement}
Not applicable.

\section*{Data Availability Statement}
Not applicable.

\section*{Conflicts of Interest}
The author declares no conflicts of interest.

\newpage

\appendix

\section{Elementwise derivations using matrix entries}

For clarity, write out the \((k,\ell)\) entry of the covariance estimator from \eqref{eq:Sigma-bariance}:
\begin{equation}
\label{eq:Sigma-bariance}
\Sigma_{k\ell}
=\frac{1}{n(n-1)}\left(
  n\sum_{i=1}^n x_{ik}x_{i\ell}
  -
  \Big(\sum_{i=1}^n x_{ik}\Big)
  \Big(\sum_{i=1}^n x_{i\ell}\Big)
\right).
\end{equation}
The centered definition is
\begin{equation}
\label{eq:Sigma-centered}
\frac{1}{n-1}\sum_{i=1}^n (x_{ik}-\bar{x}_k)(x_{i\ell}-\bar{x}_\ell)
=\frac{1}{n-1}\left(
\sum_{i=1}^n x_{ik}x_{i\ell}
-\bar{x}_k \sum_{i=1}^n x_{i\ell}
-\bar{x}_\ell \sum_{i=1}^n x_{ik}
+n\,\bar{x}_k \bar{x}_\ell
\right),
\end{equation}
where \(\bar{x}_k = (\sum_i x_{ik})/n\) and \(\bar{x}_\ell = (\sum_i x_{i\ell})/n\).

\paragraph{Elementwise derivation.}
Substitute \(\bar{x}_k\) and \(\bar{x}_\ell\) into \eqref{eq:Sigma-centered}:
\[
\frac{1}{n-1}\left(
\sum_{i=1}^n x_{ik}x_{i\ell}
-\frac{1}{n}\Big(\sum_{i=1}^n x_{ik}\Big)\Big(\sum_{i=1}^n x_{i\ell}\Big)
-\frac{1}{n}\Big(\sum_{i=1}^n x_{i\ell}\Big)\Big(\sum_{i=1}^n x_{ik}\Big)
+\frac{n}{n^2}\Big(\sum_{i=1}^n x_{ik}\Big)\Big(\sum_{i=1}^n x_{i\ell}\Big)
\right).
\]
Combine and simplify:
\[
\frac{1}{n-1}\left(
\sum_{i=1}^n x_{ik}x_{i\ell}
-\frac{1}{n}\Big(\sum_{i=1}^n x_{ik}\Big)\Big(\sum_{i=1}^n x_{i\ell}\Big)
\right)
=\frac{1}{n(n-1)}\left(
n\sum_{i=1}^n x_{ik}x_{i\ell}
-\Big(\sum_{i=1}^n x_{ik}\Big)\Big(\sum_{i=1}^n x_{i\ell}\Big)
\right),
\]
matching \eqref{eq:Sigma-bariance}. This completes the elementwise equivalence.

\paragraph{Matrix form.}
Let \(\mathbf{X} \in \mathbb{R}^{n\times p}\), \(\mathbf{s} = \mathbf{X}^\top \mathbf{1}_n\), and \(\mathbf{H} = \mathbf{I}_n - \tfrac{1}{n}\mathbf{1}_n\mathbf{1}_n^\top\).  
The covariance matrix is
\begin{equation}
\label{eq:Sigma-matrix}
\widehat{\boldsymbol{\Sigma}}
=\frac{1}{n-1}\,\mathbf{X}^\top \mathbf{H} \mathbf{X}
=\frac{1}{n-1}\left(\mathbf{X}^\top \mathbf{X} - \frac{1}{n}\mathbf{s}\,\mathbf{s}^\top\right).
\end{equation}
The \((k,\ell)\) entry of \eqref{eq:Sigma-matrix} equals \eqref{eq:Sigma-bariance}, so the matrix and scalar derivations coincide.

\paragraph{Pairwise difference identity.}
Starting from
\[
\widehat{\Sigma}_{k\ell}
=\frac{1}{n(n-1)}\sum_{1\le i<j\le n}(x_{ik}-x_{jk})(x_{i\ell}-x_{j\ell}),
\]
use
\[
\sum_{i<j} a_i a_j = \tfrac{1}{2}\!\left[\!\left(\sum_i a_i\right)^2 - \sum_i a_i^2\!\right],
\]
on columns \(k\) and \(\ell\). After expansion and cancellation one obtains \eqref{eq:Sigma-bariance}, and from \eqref{eq:Sigma-matrix} the compact form
\[
\widehat{\boldsymbol{\Sigma}}=\frac{1}{n-1}\left(\mathbf{X}^\top \mathbf{X} - \frac{1}{n}\mathbf{s}\,\mathbf{s}^\top\right).
\]

\paragraph{Computational note.}
Equation \eqref{eq:Sigma-matrix} requires one Gram product \(\mathbf{X}^\top \mathbf{X}\) and one outer product \(\mathbf{s}\mathbf{s}^\top\).  
This avoids constructing the centered matrix \(\mathbf{H}\mathbf{X}\) and keeps the cost to a single \(p\times p\) multiplication and one subtraction.  
Fast Gram routines such as RXTX~\cite{rybin2025xxt} for \(\mathbf{X}\mathbf{X}^\top\) can further reduce runtime in high-dimensional cases.

\section{Python and R references for replication}

\paragraph{Python.}
A minimal implementation (MVP) of the optimized covariance estimator is:
\begin{verbatim}
cov_bar(X) = (n * (X.T @ X) - np.outer(s, s)) / (n * (n - 1))
\end{verbatim}
with \( s = X^\top \mathbf{1}_n \) and \( n = X.\texttt{shape[0]} \).
This path triggers one level-3 BLAS multiply for \(X^\top X\) (GEMM/DSYRK) and one rank-1 update for \(s\,s^\top\) (GER), avoiding materialization of \(X-\mathbf{1}_n\bar x^\top\).
In many Python builds NumPy is linked to a generic or single-thread OpenBLAS; users often observe that raw \texttt{@} calls are fast when BLAS is available, while higher-level helpers incur extra allocations and passes \cite{so37184618,so56057835}.
Identifying the active BLAS and confirming that \texttt{np.dot}/\texttt{@} dispatch into it is standard practice \cite{so37184618}.
Under these conditions the closed form can outpace \texttt{numpy.cov}, which centers then multiplies and moves an \(n\times p\) temporary.

\paragraph{R (base).}
An (at least numerically) equivalent base-R implementation is:
\begin{verbatim}
bariance_cov <- function(X) {
  n <- nrow(X)
  s <- colSums(X)
  G <- crossprod(X)
  (n * G - tcrossprod(s, s)) / (n * (n - 1))
}
\end{verbatim}
In R, \texttt{crossprod}/\texttt{tcrossprod} map directly to level-3 BLAS (e.g., DSYRK/DGEMM) and are already the preferred fast path \cite{so15162741,so39532187}.
Guidance on BLAS-backed functions in base R points the same way \cite{so47440244}.
Base \texttt{cov()} is a compiled routine that performs centering and Gram products through BLAS, so its inner loops remain inside C/Fortran once data pointers are set.
With a multithreaded vendor BLAS, this often beats an R-level function wrapper, even if that wrapper calls \texttt{crossprod} internally.

\paragraph{Outcome.}
Python: the closed form is faster when \texttt{numpy.cov} does extra allocations and the build is not strongly tuned; the single GEMM\(+\)GER path wins \cite{so37184618,so56057835}.
R: \texttt{cov()} already leverages BLAS through compiled code; \texttt{cov()} remains faster than an R-level wrapper that still returns to the interpreter between BLAS calls \cite{so15162741,so39532187,so47440244}.


\newpage

\begin{thebibliography}{99}

\bibitem{anderson2003introduction}
ANDERSON, T.~W.  
\emph{An Introduction to Multivariate Statistical Analysis}.  
Wiley, 3rd~ed., 2003.

\bibitem{arellano1987computing}
ARELLANO, M.  
Computing robust standard errors for within-groups estimators.  
\emph{Oxford Bulletin of Economics and Statistics}, \textbf{49}(4):431--434, 1987.

\bibitem{benisrael2003generalized}
BEN-ISRAEL, A. and GREVILLE, T.~N.~E.  
\emph{Generalized Inverses: Theory and Applications}.  
Springer, 2nd~ed., 2003.

\bibitem{cameron2006robust}
CAMERON, A.~C., GELBACH, J.~B., and MILLER, D.~L.  
Robust inference with multi-way clustering.  
\emph{NBER Technical Working Paper No.~327}, 2006.

\bibitem{cochran1977sampling}
COCHRAN, W.~G.  
\emph{Sampling Techniques}.  
Wiley, 3rd~ed., 1977.

\bibitem{demmel1997applied}
DEMMEL, J.~W.  
\emph{Applied Numerical Linear Algebra}.  
SIAM, 1997.

\bibitem{golub2013matrix}
GOLUB, G.~H. and VAN~LOAN, C.~F.  
\emph{Matrix Computations}.  
Johns Hopkins University Press, 4th~ed., 2013.

\bibitem{hansen1982large}
HANSEN, L.~P.  
Large sample properties of generalized method of moments estimators.  
\emph{Econometrica}, \textbf{50}(4):1029--1054, 1982.

\bibitem{harville1997matrix}
HARVILLE, D.~A.  
\emph{Matrix Algebra from a Statistician’s Perspective}.  
Springer, 1997.

\bibitem{higham2002accuracy}
HIGHAM, N.~J.  
\emph{Accuracy and Stability of Numerical Algorithms}.  
SIAM, 2nd~ed., 2002.

\bibitem{horn1985matrix}
HORN, R.~A. and JOHNSON, C.~R.  
\emph{Matrix Analysis}.  
Cambridge University Press, 1985.

\bibitem{huber1967under}
HUBER, P.~J.  
The behavior of maximum likelihood estimates under nonstandard conditions.  
In \emph{Proceedings of the Fifth Berkeley Symposium on Mathematical Statistics and Probability}, 1967.

\bibitem{magnus1988matrix}
MAGNUS, J.~R. and NEUDECKER, H.  
\emph{Matrix Differential Calculus with Applications in Statistics and Econometrics}.  
Wiley, 1988.

\bibitem{newey1987simple}
NEWEY, W.~K. and WEST, K.~D.  
A simple, positive semi-definite, heteroskedasticity and autocorrelation consistent covariance matrix.  
\emph{Econometrica}, \textbf{55}(3):703--708, 1987.

\bibitem{reichel2025bariance}
REICHEL, F.  
On Bessel’s correction: Unbiased sample variance, the bariance, and a novel runtime-optimized estimator.  
\emph{arXiv preprint arXiv:2503.22333v5}, 2025.  
Available at: \url{https://doi.org/10.48550/arXiv.2503.22333}

\bibitem{rybin2025xxt}
RYBIN, D., ZHANG, Y., and LUO, Z.-Q.  
XX$^{\mathsf{T}}$ can be faster.  
\emph{arXiv preprint arXiv:2505.09814}, 2025.  
Available at: \url{https://doi.org/10.48550/arXiv.2505.09814}

\bibitem{searle2006matrix}
SEARLE, S.~R.  
\emph{Matrix Algebra Useful for Statistics}.  
Wiley, 2006.

\bibitem{seber2003linear}
SEBER, G.~A.~F. and LEE, A.~J.  
\emph{Linear Regression Analysis}.  
Wiley, 2nd~ed., 2003.

\bibitem{so37184618}
STACK OVERFLOW COMMUNITY.  
Find out if / which BLAS library is used by NumPy.  
Available at: \url{https://stackoverflow.com/questions/37184618/find-out-if-which-blas-library-is-used-by-numpy}  
(Accessed: 6~Nov~2025).

\bibitem{so56057835}
STACK OVERFLOW COMMUNITY.  
How to decrease the output time of covariance function.  
Available at: \url{https://stackoverflow.com/questions/56057835/how-to-decrease-the-output-time-of-covariance-function}  
(Accessed: 6~Nov~2025).

\bibitem{so15162741}
STACK OVERFLOW COMMUNITY.  
What is R’s \texttt{crossprod} function?  
Available at: \url{https://stackoverflow.com/questions/15162741/what-is-rs-crossproduct-function}  
(Accessed: 6~Nov~2025).

\bibitem{so39532187}
STACK OVERFLOW COMMUNITY.  
Any faster R function than \texttt{tcrossprod} for symmetric dense matrix multiplication?  
Available at: \url{https://stackoverflow.com/questions/39532187/r-any-faster-r-function-than-tcrossprod-for-symmetric-dense-matrix-multiplica}  
(Accessed: 6~Nov~2025).

\bibitem{so47440244}
STACK OVERFLOW COMMUNITY.  
Which R functions are based exactly on BLAS, ATLAS, LAPACK, and so on?  
Available at: \url{https://stackoverflow.com/questions/47440244/which-r-functions-are-based-exactly-on-blas-atlas-lapack-and-so-on}  
(Accessed: 6~Nov~2025).

\bibitem{tsay2013multivariate}
TSAY, R.~S.  
\emph{Multivariate Time Series Analysis: With R and Financial Applications}.  
Wiley, 2013.

\bibitem{white1980heteroskedasticity}
WHITE, H.  
A heteroskedasticity-consistent covariance matrix estimator and a direct test for heteroskedasticity.  
\emph{Econometrica}, \textbf{48}(4):817--838, 1980.

\bibitem{wooldridge2010econometric}
WOOLDRIDGE, J.~M.  
\emph{Econometric Analysis of Cross Section and Panel Data}.  
MIT Press, 2nd~ed., 2010.

\end{thebibliography}
\end{document}